# Stark Decline in Journalists' Use of Preprints Post-pandemic

Juan Pablo Alperin[1,*], Kenneth Shores[2], Alice Fleerackers[1,3], Natascha Chtena[1]

Author Affiliations
[1]*School of Publishing, Simon Fraser University, 515 West Hastings Street, Vancouver, BC V6B 5K3, Canada*
[2]*School of Education, University of Delaware, 113 Willard Hall Education Building, Newark, DE 19716, USA*
[3]*School of Journalism, Writing, and Media, University of British Columbia, 6388 Crescent Road Vancouver, BC V6T 1Z2, Canada*

[*] Corresponding Author: juan@alperin.ca

# Stark Decline in Journalists' Use of Preprints Post-pandemic

## Abstract

The COVID-19 pandemic accelerated the use of preprints, aiding rapid research dissemination but also facilitating the spread of misinformation. This study analyzes media coverage of preprints from 2014 to 2023, revealing a significant post-pandemic decline. Our findings suggest that heightened awareness of the risks associated with preprints has led to more cautious media practices. While the decline in preprint coverage may mitigate concerns about premature media exposure, it also raises questions about the future role of preprints in science communication, especially during emergencies. Balanced policies based on up-to-date evidence are needed to address this shift.

# Introduction

Preprints offer an exciting opportunity to sidestep many of the challenges of traditional academic publishing—namely, they offer rapid dissemination of research and lower barriers for readers and authors. However, they also present a risk—that the dissemination of unreviewed science could lead to the spread of harmful mis- and disinformation. The COVID-19 pandemic vividly underscored both sides of preprints.

Preprints were a driving force in the pandemic response, with both researchers and journalists using them to share emerging evidence at levels not seen before (Fleerackers et al., 2024; Fraser et al., 2021). Much of the media coverage of COVID-19 preprints was helpful, with some of the most widely reported preprints covering topics such as COVID-19's aerosol and surface stability and the effectiveness of social distancing and other interventions (Fraser et al., 2021). The rise in preprints narrowed the gap between researchers, policymakers, and the public, providing unhampered access to research material with direct social relevance when it was sorely needed.

Yet, a handful of problematic preprints posted during the pandemic fueled misinformation and conspiracy theories about the SARS-CoV-2 virus. One such preprint—posted on bioRxiv on January 31, 2020, and withdrawn only two days later—claimed to have identified similarities between the DNA of the novel coronavirus and that of HIV and suggested that COVID-19 might therefore have been engineered by humans (Davey, 2020). Although the academic community was quick to challenge these claims, the findings had already been widely disseminated in mainstream and alternative media, as well as on discussion boards, in podcasts, and on social media.

The misinformation that spread through these high-profile cases brought new urgency to concerns about the public risks of preprints. Scientists were urged to "be extremely cautious about releasing preprint results" (Roy & Edwards, 2022, p. 1), while journalists were told to "take special precautions" (Khamsi, 2020) in vetting their quality. There was a growing consensus that the scholarly community and the media needed to "do better" in communicating about these unreviewed studies with the public (Caulfield et al., 2021).

Previous research shows that media coverage of COVID-related preprints surged during the early pandemic, while coverage of preprints on other topics saw a slight decline (Fleerackers et al., 2024). However, now that the emergency phase of the pandemic has passed, it is uncertain to what extent journalists will continue to use preprints. We suggest three hypotheses: 1) that the pandemic taught journalists about the value of preprints and that this resulted in a modest increase in preprint use moving forward; 2) that COVID was unique and that preprint use returned to pre-COVID use rates; and 3) that preprint use during the pandemic chastened journalists and alarmed them to the dangers of unvetted scholarship, resulting in a decrease in preprint use.

The present study tests these hypotheses by analyzing the number of mentions of preprints in the media between January 1, 2014, and December 31, 2023, relative to the mentions of research published in journals indexed in the Web of Science (WoS).

## Materials and Methods

This study utilizes data from Altmetric collected in January 2024 and augments it with article and preprint metadata from Crossref and arXiv. We focused on the 92 media outlets that had mentioned WoS research at least 100 times every year between 2014 and 2023 (cf. 2). After filtering and cleaning, our final preprint sample comprised 61,115 mentions of 21,490 preprints in 48,472 stories published by the 92 outlets in our sample (a preprint could be covered in several stories and a single story could mention multiple preprints). The final WoS-indexed research sample comprised 2,325,207 mentions of 632,815 research outputs across 1,542,571 stories published by the same 92 outlets.

To analyze the proportion of research-based media coverage that mentions preprints, we estimated multivariate fractional regressions. We estimated counts of preprints and counts of WoS publications mentioned by the media as Poisson regressions. Further details of our methods can be found in the detailed methods section below.

## Results and Conclusion

Our analysis provides support for the third hypothesis: following the end of the COVID-19 pandemic, media coverage of preprints has declined well below rates forecasted by pre-pandemic trends. This decline can be seen both in the proportion of research-based media coverage that mentions preprints (Fig. 1) and in absolute numbers (Fig. 2), following the initial surge during the pandemic's early stages. We calculate that, one year into the pandemic, the share of research-based media coverage mentioning preprints had increased by 2.1 percentage points (PP) ($p<0.000$; 95% CI [2.6 – 1.6]; predicted share of preprint mentions on January 10, 2021 based on pre-pandemic trends equal to 2.4 PP; observed share of preprint mentions equal to 4.5 PP. By the end of 2023, it had decreased by 1.4 PP ($p<0.000$; 95% CI [0.9 – 2.0]; predicted share of preprint mentions on December 31, 2023 based on pre-pandemic trends equal to 3.5 PP; observed share of preprint mentions equal to 2.1 PP) (Fig. 1). Importantly, these changes in the proportion of research-based media coverage mentioning preprints were not caused by changes in journalists' treatment of WoS publications, as media coverage of WoS publications throughout this period was largely invariant to the pandemic (Fig. 2).

Across the nine servers included in our analysis, by December 2023, we estimate a 68% decline in the number of preprints mentioned in the media ($p<0.000$; 95% CI [0.58 – 0.78]), relative to what we would anticipate based on pre-pandemic trends (Table 1; predicted preprint mentions based on pre-pandemic trends equal to 1473.6; observed preprint mentions equal to 473.6). The size of the decline differs by server, ranging from 77% (bioRxiv) to as high as 94% (SSRN). Only media coverage of arXiv preprints remains unaffected, with estimated mentions of preprints at the same level as predicted using pre-pandemic data. Moreover, declines in media coverage of preprints remain the same or become more pronounced when controlling for the number of preprints posted to the servers. Indeed, while the number of available preprints increased steadily and linearly throughout and after the pandemic, media mentions of preprints increased dramatically during COVID-19 and then declined.

**Figure 1.** Proportion of research-based media coverage that mentions preprints per day.

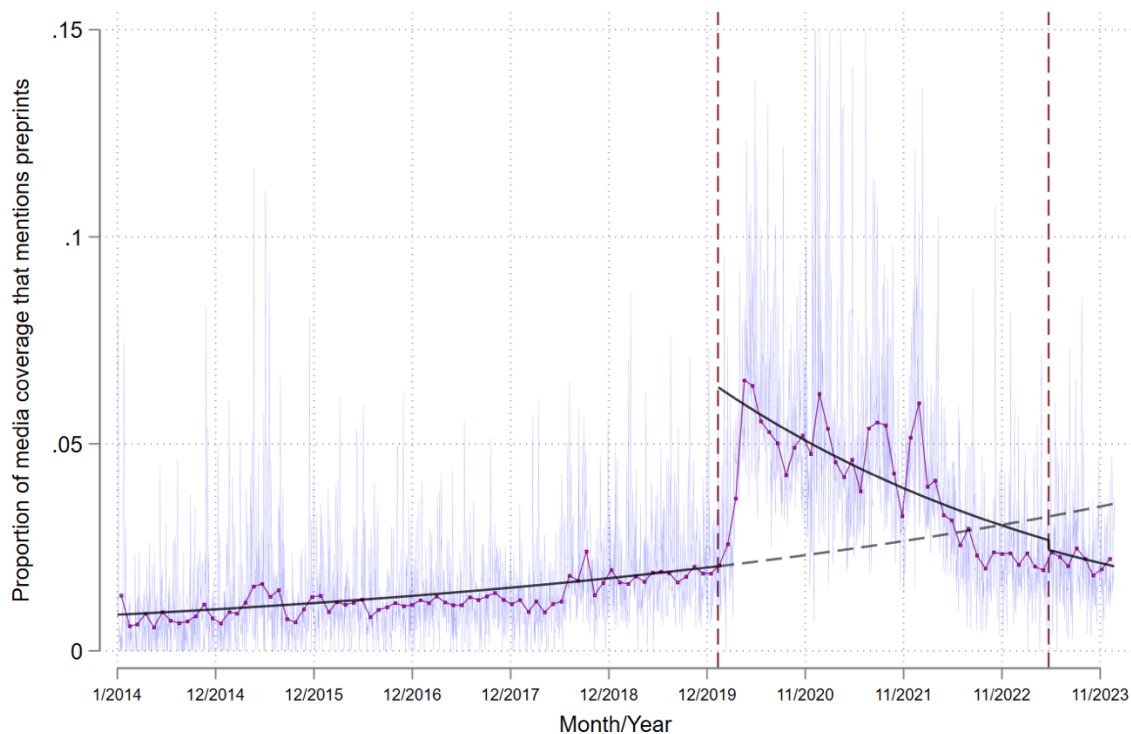

Note: Figure shows the proportion of research-based media coverage that mentions preprints per day beginning January 1, 2014, and ending December 31, 2023. The thin blue line plots the daily proportion of research-based media coverage that mentions preprints (i.e., mentions of preprints relative to mentions of preprints and WoS publications). The thicker purple line with marker symbols plots the average monthly proportion of research-based media coverage that mentions preprints, with each marker symbol plotted on the 15$^{th}$ day of each month. The solid gray line is the predicted proportion of research-based media coverage that mentions preprints based on a fractional logistic regression that models time (in days) linearly, an intercept shift for when the COVID-19 pandemic begins, an interaction between linear time and the pandemic era, and an intercept shift for when the pandemic ends. The dashed gray line represents the predicted proportion of research-based media coverage that mentions preprints based on the trend in preprint mentions prior to the onset of the COVID-19 pandemic. Vertical lines mark the start and end of the COVID-19 pandemic according to the World Health Organization.

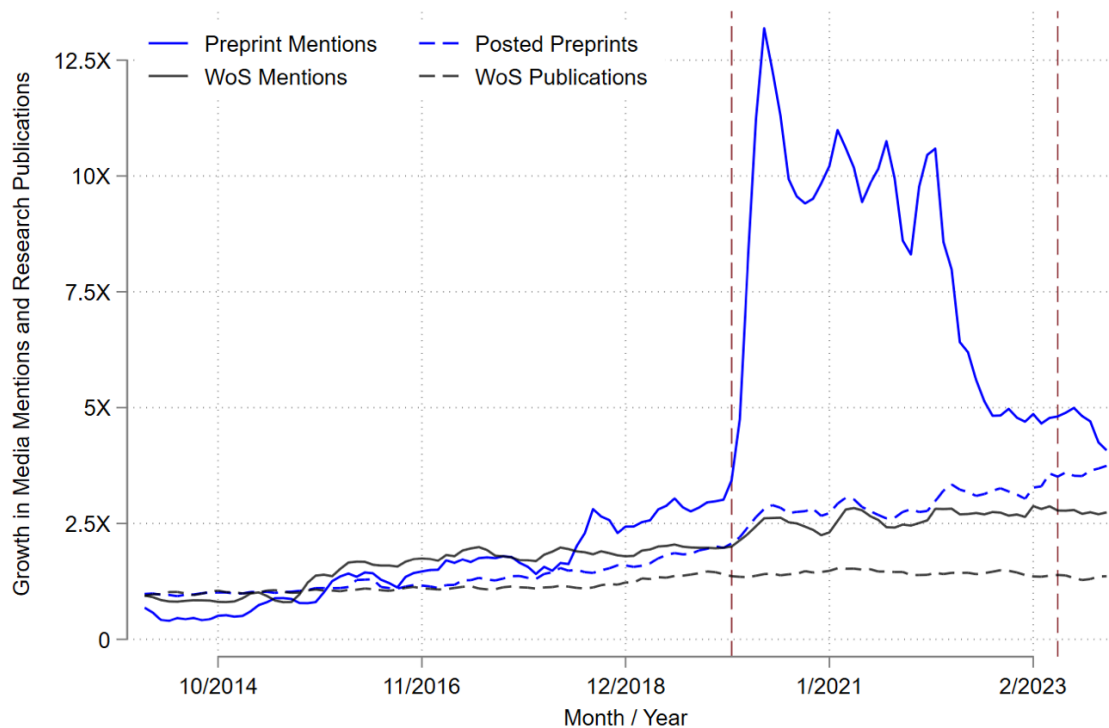

**Figure 2.** Growth rates of total preprint mentions, total preprints posted, total WoS mentions, and total WoS publications.

Note: Growth rates are calculated from January of 2014 for all nine preprint servers and for the WoS. Total number of preprint mentions, preprints posted, WoS mentions, and WoS publications are calculated for each year-month; the growth rate is the ratio of the current count in the year-month relative to the count in January 2014. Vertical lines mark the start and end of the COVID-19 pandemic according to the World Health Organization.

**Table 1**. Proportional Decline in Counts of Media Mentions (One minus the Ratio of Predicted Counts to Observed Media Mentions)

|  | [1] Unconditional |  | [2] Conditional |  |
|---|---|---|---|---|
| All Servers | 0.68 | *** | 0.71 | *** |
|  | (0.05) |  | (0.08) |  |
| WoS | 0.35 | *** | 0.35 | *** |
|  | (0.04) |  | (0.04) |  |
| bioRxiv/medRxiv | 0.77 | ** | 0.83 | ** |
|  | (0.09) |  | (0.09) |  |
| bioRxiv | 0.86 | *** | 0.88 | * |
|  | (0.05) |  | (0.06) |  |
| arXiv | 0.00 | *** | 0.04 | *** |
|  | (0.24) |  | (0.24) |  |
| NBER | 0.89 |  | 0.89 |  |
|  | (0.08) |  | (0.08) |  |
| SSRN | 0.94 | *** | 0.96 | *** |
|  | (0.01) |  | (0.01) |  |

Notes: This table shows the ratio of the predicted number of preprint mentions based on pre-pandemic trends relative to the estimated number of preprint mentions in December, 2023. Model coefficients are estimated using Poisson regression, and the predicted ratio of preprint mentions after the pandemic relative to predicted preprint mentions using pre-pandemic trends is generated from model coefficients. Model [1] Unconditional reports the estimated ratio without additional covariates, and Model [2] Conditional reports the estimated ratio controlling for the number of preprints posted and WoS publications each month. For the WoS result we also control for a January-specific effect due to publications without specific publication dates being assigned to January 1 by the WoS indexing process. For the medRxiv result, we combine medRxiv mentions with bioRxiv mentions because there is insufficient medRxiv data to calculate pre-COVID trends.

Notably, we also estimate a 35% decline in the number of WoS publications mentioned in the media post-pandemic relative to pre-pandemic trends. As with preprints, this decline remains unchanged when controlling for the number of WoS publications. However, although this decline relative to the expected is substantial, it pales in comparison to the decline in media mentions of preprints, especially for the two social sciences servers (SSRN and NBER).

This decline in preprint coverage supports findings from interviews we conducted with health and science journalists in prior work (Fleerackers et al., 2022). These specialized journalists voiced concerns about the potential risks of spreading misinformation and ultimately harming their audiences. While they found vetting preprints challenging, many reported using rigorous strategies to do so: triangulating findings with those of other studies and seeking critical input from unaffiliated scientists. Most importantly, these journalists strove to put the interests of the public first, limiting their coverage of preprints to instances when they felt the potential benefits of doing so outweighed the possible risks. Combined with the results of the current analysis, there is reason to believe that other journalists may also be well-aware of the risks of covering preprints and are now taking a more cautious approach to when and how they report on them.

This reduction in media coverage may alleviate one of the chief concerns expressed in a recent global survey of attitudes towards preprinting: "premature media coverage" (Ni & Waltman, 2024). Given the timing of the survey (late 2022 and early 2023), it is likely that this concern, expressed most clearly by those in life and health sciences, was in no small part informed by the explosion of preprints that circulated in the media during the pandemic—particularly preprints that were later found to be flawed or problematic. However, our findings suggest that journalists have become more, not less, discerning about when they cover these unreviewed studies.

This change in journalists' knowledge, attitudes, or willingness to cover preprints may alleviate concerns about the spread of misinformation, but also raises new questions that warrant reflection. Does the change in journalists' willingness to cover preprints signal a broader transformation in their understanding or attitudes towards science and, particularly, science-in-progress? Conversely, how might the public's knowledge and attitudes towards preprints, research, or to science-based journalism, change in the aftermath of the pandemic? And, perhaps most consequentially, when another emergency presents itself, will journalists, scholars, and other science communicators be able, or willing, to use preprints to rapidly circulate life-saving research in the same way?

As we set policies, develop guidelines, and establish best practices for how and when to use preprints going forward, it is essential that we recognize that journalists' use of COVID-19 preprints was—like so many aspects of the pandemic—a trial by fire and a crucible of learning. The concerns evidenced during the pandemic period should not make us forget that it was rapid and low-barrier publishing, informal review (not formal peer review), and media coverage that made preprints so effective in a time of crisis.

# Detailed Methods

**I. Data Collection**

We collected mentions of research in the news media using Altmetric, a company tracking mentions of research outputs across various digital media platforms. The Altmetric "Mainstream Media" category has been shown to be reliable for identifying research mentions across a predetermined list of English-language media sources (1). As such, we used a list of 94 English-language outlets curated for an earlier study (2) that had mentioned research at least 100 times every year between 2014 and 2020. To continue with the same criteria, we removed two outlets that did not meet this 100-article threshold between 2021 and 2023, leaving us with a list of 92 outlets covering a wide range of topics (e.g., science/technology, health/medicine, business, general news, etc). By querying the Altmetric Explorer API in the first week of 2024, we downloaded details of every time any research (including preprints and Web of Science [WoS] research outputs) was mentioned in these outlets.

We subsequently queried the Crossref API to collect metadata details for every DOI that had been mentioned in the 92 media outlets and the arXiv API for every arXiv ID mentioned. Using this metadata, we identified which mentions belonged to the following preprint servers: arXiv, bioRxiv, chemRxiv, medRxiv, NBER, OSF, Research Square, SSRN, and techRxiv. We compared the ISSN found in the Crossref metadata with a list of the ISSNs of every journal in the WoS to identify which mentions were associated with journals indexed in that database. For simplicity, we refer to these documents published in journals that are indexed in the Web of Science as "WoS publications" and to mentions of these documents as "WoS mentions."

Following the same approach as Fleerackers et al. (2024), we determined the most likely publication date for each preprint or WoS publication. For arXiv, it was the date provided by the arXiv API; for SSRN, it was either the "first posted on" date provided by Altmetric or Crossref's DOI creation date, whichever came first, and for all other servers, it was the DOI creation date.

Finally, between February and April 2024, we collected the number of preprints posted each month for all the servers in our dataset. We collected the counts for arXiv, bioRxiv, chemRxiv, medRxiv, OSF, Research Square, and techRxiv by querying the Dimensions database. The number of preprints posted on NBER each month were not available from Dimensions, so we downloaded the metadata for every preprint in the IDEAS/RePEc repository and calculated counts based on their publication dates. Because preprint counts for SSRN reported in Dimensions showed large volatility, we downloaded monthly preprint counts by scraping reports made available on the SSRN homepage. Finally, we collected the number of documents found in WoS by querying the WoS using the web interface and filtering for one month at a time using the date of publication (DOP) filter.

**II. Data Cleaning**

We identified 71,891 mentions of preprints across the 92 outlets. From these, we filtered out 8,081 mentions of preprints with publication dates before 2013, another 641 mentions of preprints whose publication dates suggested they were postprints rather than preprints (i.e., the preprint date was less than a week before, or was after, the corresponding paper's publication date), and an additional 327 mentions where the preprint publication date was after the date of the mention (i.e., could not have been possible and must therefore be caused by a metadata error). Finally, we removed 1,727 duplicate mentions. In total, filtering led to the exclusion of 10,776 mentions (15.0% of original dataset). Our final preprint sample comprised 61,115 mentions of 21,490 preprints in 48,472 stories published by the 92 outlets in our sample.

We identified 2,708,008 WoS mentions across the 92 outlets. From these, we filtered out 350,884 mentions with publication dates before 2013 and removed 31,917 duplicate mentions. In total, filtering led to the exclusion of 382,801 mentions (14.1% of original dataset). The final WoS sample comprised 2,325,207 mentions of 632,815 distinct research outputs across 1,542,571 stories published by the 92 outlets in our sample.

**III. Statistical Methods**

We employ a fractional logistic regression model to examine the proportion of research-based media coverage that mentions preprints (i.e., the number of mentions of preprints relative to mentions of preprints and WoS outputs; *dailyshare*). The dependent variable, *dailyshare*, is bounded between 0 and 1 and represents the proportion of media mentions of research that are from preprint servers on day *t*. The primary explanatory variables include the days where media mentions are tabulated beginning on January 1, 2014 and ending December 31, 2023 (*groupdaily*), the days where media mentions are tabulated beginning on Jan 10, 2020 when COVID-19 was declared a pandemic by the World Health Organization (*timesince*), an indicator variable for the start of the COVID-19 pandemic on that date (*covidstarts*), an indicator variable for the end of the COVID-19 pandemic (*covidends*), and an interaction between *timesince* and *covidstarts*. Because the time-period after *covidends* is short we do not include a separate interaction between *groupdaily* and this variable. The fractional logistic regression model estimates the log-odds of the dependent variable, which is then transformed to probabilities using the logistic function. Statistical analysis was performed using Stata version 17 (StataCorp, 2021).

The fractional logistic regression model can be specified as follows:

$$logit(dailyshare)_t = \beta_0 + \beta_1 Groupdaily + \beta_2 Covidstarts + \beta_3 Timesince * Covidstarts + \beta_4 Covidends + \varepsilon_t$$

where:

- $logit(dailyshare)_{it}$ is the log-odds of the proportion of research-based media coverage that mentions preprints for day *t*.
- $\beta_0$ is the intercept.
- $\beta_1$ is the coefficient for the number of daily media mentions.
- $\beta_2$ is the coefficient for the start of the COVID-19 pandemic.
- $\beta_3$ is the coefficient for the interaction term between daily media mentions beginning with the COVID-19 pandemic and the start of the COVID-19 pandemic.
- $\beta_4$ is the coefficient for the end of the COVID-19 pandemic.

- $\varepsilon_t$ is the error term and is robust to heteroskedasticity.

We then calculate the marginal change in predicted proportions of research-based media coverage that mentions preprints one year into the COVID-19 pandemic (January 10, 2021) due to the COVID-19 effect, i.e., the average change in predicted proportions of media mentions focused on preprints one-year into the pandemic. We then calculate the predicted change in the proportion of research-based media coverage that mentions preprints after the pandemic relative to the predicted proportion of research-based media coverage that mentions preprints based on pre-pandemic trends. We conduct these tests by transforming model coefficients from log-odds to predicted proportions and plugging in known values of the regressors.

Specifically, we conducted a post-estimation test using the nonlinear combination of estimators (nlcom) command in Stata. This test compared the predicted counts for two scenarios:

1. **Scenario 1**: Predicted count of mentions on January 10, 2021, based solely on pre-pandemic trends (i.e., setting COVID-19-related variables to zero).
2. **Scenario 2**: Predicted count of mentions on January 10, 2021, one year into the pandemic, including the effects of all variables, especially the main effect of $\beta_2$, which represents the change in the share of preprints mentioned by the media when the pandemic began.

These scenarios can then be estimated as the predicted proportion of research-based media coverage that mentions preprints one-year into the pandemic using pre-pandemic trends as:

$\hat{P}_1 = \frac{1}{1+\exp{(-(\beta_0+\beta_1*2{,}566))}}$, where 2,566 corresponds to the date January 10, 2021 in our sample

$\hat{P}_2 = \frac{1}{1+\exp{(-(\beta_0+\beta_1*2{,}566+\beta_2+\beta_3*366))}}$, which captures the predicted change in the share of preprint media mentions on January 10, 2021 and includes the estimated pandemic effect and the linear change in preprint mentions 366 days into the COVID-19 pandemic. The estimated change in preprint mentions is the difference between $\hat{P}_1$ and $\hat{P}_2$, which we estimated using the nlcom command in Stata.

We conduct a similar test to estimate the change in share of preprint mentions after the pandemic, substituting 3,630 for 2,566 and 1,430 for 366, which corresponds to the date December 10, 2023 in our sample, and including the effect of $\beta_4$ when calculating $\hat{P}_2$.

**Figure 2.**

Figure 2 is based on a collapsed dataset of counts of preprint mentions, posted preprints, WoS mentions, and WoS publications. This collapsed data is available at the server level (e.g., with counts for NBER and arXiv individuated) and pooled across all servers. Figure 1 represents the growth rate in mentions and publications for all preprint servers and WoS separately. The growth rate is calculated as the observed mention/publication count relative to the mention/publication count in January 2014, the month-year our data begins. WoS publications are sometimes reported annually and assigned a January publication date, resulting in an apparent surge in publications in January of each year. To allow for annual secular growth in publications but remove the January-specific effect, we calculate the average annual publication count for WoS excluding January and then take the difference between the January publication count and the average annual publication count minus January. This difference represents the "January effect"; we then remove the January effect from January and divide that effect by 12, distributing it across all months in the year. Lastly, we apply a simple moving average smoother that replaced the observed mention/publication with the average of the preceding, observed, and subsequent month (i.e., a +/-1 smoother). No other modifications to the data were made.

**Table 1.**

To generate the reported ratios of estimated preprint mentions relative to predicted preprint mentions based on pre-pandemic trends, we employed a Poisson regression model to estimate the number of media mentions in a given month-year. The model accounts for various factors, including the introduction of the medRxiv server, the start and end of the COVID-19 pandemic, and two time trends, one for the pre-pandemic period and another for the pandemic period. We do not include an interaction between the post-pandemic period and months because the time-period is so short. When we estimate counts for all servers, we collapse the data to obtain the total count of preprint research mentioned by the media; for individual servers, we use the server-specific count. The model is specified as follows:

Equation (2)

$$\log(\lambda_{it}) = \beta_0 + \beta_1 Medrxiv + \beta_2 TimeCent + \beta_3 CovidStarts + \beta_4 TimeSince + \beta_5 CovidEnds + \varepsilon_{it}$$

Where:

- $\lambda_{it}$ is the expected count of media mentions for preprint/publication *i* posted on month *t*.
- $\beta_0$ is the intercept.
- $\beta_1$ represents the effect of the medRxiv server coming online (*Medrxiv* is a binary variable).
- $\beta_2$ represents the effect of time since the start of the pandemic (*TimeCent* is a continuous variable centered at the start of the pandemic).
- $\beta_3$ represents the effect of the COVID-19 pandemic starting (*CovidStarts* is a binary variable).
- $\beta_4$ represents the effect of time since the start of the COVID-19 pandemic (*TimeSince* is a continuous variable coded as 0 prior to the pandemic and then set equal to 1 and increasing incrementally by 1 for each month beginning with the start of the COVID-19 pandemic).
- $\beta_5$ represents the effect of the COVID-19 pandemic ending (*CovidEnds* is a binary variable).
- $\varepsilon_{it}$ is the error term and is robust to heteroskedasticity.

Following the estimation of the Poisson regression model, we conducted a post-estimation test using the nonlinear combination of estimators (nlcom) command in Stata. This test compared the predicted counts for two scenarios:

1. **Scenario 1**: Predicted count of mentions 48 months after the start of the pandemic, which corresponds to December 2023, including the effects of all variables.
2. **Scenario 2**: Predicted count of mentions 48 months after the start of the pandemic, based solely on pre-pandemic trends (i.e., setting COVID-19-related variables to zero).

Let $\lambda_1$ represent the predicted count of mentions for Scenario 1 and $\lambda_2$ represent the predicted count of mentions for Scenario 2.

Predicted Count for Scenario 1:

$\log(\lambda_{it}) = \beta_0 + \beta_1 + \beta_2 * 48 + \beta_3 + \beta_4 * 47 + \beta_5$; where 47 represents the number of months after the first month of the COVID-19 pandemic.

Predicted Count for Scenario 2:

$\log(\lambda_{it}) = \beta_0 + \beta_1 + \beta_2 * 48$

The nonlinear combination of these estimates provides the ratio of the predicted counts; we subtract this ratio from 1 to represent the decline in preprint mentions:

$$\frac{\widehat{\lambda_1}}{\widehat{\lambda_2}} = 1 - \left[\frac{\exp(\beta_0 + \beta_1 Medrxiv + \beta_2 TimeCent * 48 + \beta_3 CovidStarts + \beta_4 TimeSince * 47 + \beta_5 CovidEnds)}{\exp(\beta_0 + \beta_1 Medrxiv + \beta_2 TimeCent * 48)}\right]$$

This equation then then simplifies to

$$\frac{\widehat{\lambda_1}}{\widehat{\lambda_2}} = 1 - exp(\beta_3 CovidStarts + \beta_4 TimeSince * 47 + \beta_5 CovidEnds)$$

In the conditional models, we also include counts of preprints/publications (total across servers, for WoS, and individually by preprint server, in separate regressions) as control variables. For regressions for WoS, we include an additional binary variable for the month of January to account for the fact that, for journals that do not provide month of publication, the WoS indexing approach assign the entire year's publications to January.

## Acknowledgments

The authors wish to thank Altmetric for providing this study's data free of charge for research purposes. We would also like to thank Dr. Diego Chavarro and Mr. Eric Schares for their assistance in collecting counts of the number of preprints published each month. This work was supported by the Social Science and Humanities Research Council (SSHRC, Canada), Grant #2005-2021-0011.

## References

Caulfield, T., Bubela, T., Kimmelman, J., & Ravitsky, V. (2021). Let's do better: Public representations of COVID-19 science. *FACETS*, *6*, 403–423. https://doi.org/10.1139/facets-2021-0018

Davey, M. (2020, May 29). Covid-19 study on hydroxychloroquine use questioned by 120 researchers and medical professionals. *The Guardian*. https://www.theguardian.com/world/2020/may/29/covid-19-surgisphere-hydroxychloroquine-study-lancet-coronavirus-who-questioned-by-researchers-medical-professionals

Fleerackers, A., Moorhead, L., Maggio, L. A., Fagan, K., & Alperin, J. P. (2022). *Science in motion: A qualitative analysis of journalists' use and perception of preprints* (p. 2022.02.03.479041). bioRxiv. https://doi.org/10.1101/2022.02.03.479041

Fleerackers, A., Shores, K., Chtena, N., & Alperin, J. P. (2024). Unreviewed science in the news: The evolution of preprint media coverage from 2014–2021. *Quantitative Science Studies*, *5*(2), 297–316. https://doi.org/10.1162/qss_a_00282

Fraser, N., Brierley, L., Dey, G., Polka, J. K., Pálfy, M., Nanni, F., & Coates, J. A. (2021). The evolving role of preprints in the dissemination of COVID-19 research and their impact on the science communication landscape. *PLOS Biology*, *19*(4), e3000959. https://doi.org/10.1371/journal.pbio.3000959

Khamsi, R. (2020, June 1). Problems with preprints: Covering rough-draft manuscripts responsibly. *The Open Notebook*. https://www.theopennotebook.com/2020/06/01/problems-with-preprints-covering-rough-draft-manuscripts-responsibly/

Ni, R., & Waltman, L. (2024). To preprint or not to preprint: A global researcher survey. *Journal of the Association for Information Science and Technology*, *75*(6), 749–766. https://doi.org/10.1002/asi.24880

Ortega, J.L. (2020). Altmetrics data providers: A meta-analysis review of the coverage of metrics and publication. *El Profesional de La Información*, 29 . https://doi.org/10.3145/epi.2020.ene.07

Roy, S., & Edwards, M. A. (2022). Addressing the preprint dilemma. *International Journal of Hygiene and Environmental Health*, *240*, 113896. https://doi.org/10.1016/j.ijheh.2021.113896